\documentclass[reprint,onecolumn,longbibliography,notitlepage,showkeys,
superscriptaddress,
nofootinbib,
 amsmath,amssymb,
 aps,
prd,
]{revtex4-2}
\usepackage{orcidlink}
\usepackage{epsfig}

\usepackage{times}
\usepackage{hyperref}

\definecolor{C1}{RGB}{39, 242, 237} 
\definecolor{C2}{RGB}{0, 0, 255} 
\definecolor{C3}{RGB}{255, 0, 0}

\hypersetup{
    pagebackref=true,               
    hyperindex=true,                
    colorlinks=true,                
    breaklinks=true,                
    urlcolor= C2,                
    linkcolor= C3,                
    bookmarks=true,                 
    bookmarksopen=false,
    filecolor=black,
    citecolor= C2,
    linkbordercolor=blue
    }
\usepackage{graphicx}
\usepackage{orcidlink}
\usepackage{mwe}
\usepackage[T1]{fontenc}
\DeclareMathOperator{\sech}{sech}

\begin{document}
\title{Wormhole Formation with Gravitational Particle Creation Mechanism}

\author{Mrinnoy M. Gohain\orcidlink{0000-0002-1097-2124}}
\email{mrinmoygohain19@gmail.com}
\affiliation{%
 Department of Physics, Dibrugarh University, Dibrugarh \\
 Assam, India, 786004}

\author{Kalyan Bhuyan\orcidlink{0000-0002-8896-7691}}%
 \email{kalyanbhuyan@dibru.ac.in}
\affiliation{%
 Department of Physics, Dibrugarh University, Dibrugarh \\
 Assam, India, 786004}%
\affiliation{Theoretical Physics Divison, Centre for Atmospheric Studies, Dibrugarh University, Dibrugarh, Assam, India 786004}

\author{Chayanika Chetia\orcidlink{0000-0003-4307-5523}}
\email{chayanikachetia111@gmail.com}
\affiliation{%
 Department of Physics, Dibrugarh University, Dibrugarh \\
 Assam, India, 786004}

\begin{abstract}
We explore the possibility of formation of a traversable wormhole in General Relativity supported by particle creation mechanism. The repulsive back-reaction pressure generated through this mechanism can be thought of as a source of sustaining a traversable wormhole. In the first part of this paper, we model a wormhole geometry by assuming an inverse powerlaw variation of particle creation pressure within the wormhole geometry, and the shape function for the wormhole is obtained, assuming a finite redshift function. By stabilizing the wormhole structure, based on the causality of sound-speed we obtained the existence range of the parameter $\beta$ associated with particle creation mechanism. The general shape function obtained is found to adhere to the feasibility conditions of a viable wormhole geometry. Then we studied the 2D and 3D embedding derived for the obtained shape function. In the second part of the paper, we followed the reverse approach of the aforementioned treatment, where we derived the particle creation pressures by assuming three commonly utilized toy shape-functions. The energy conditions are then investigated for all these cases.  In essence, particle creation phenomena inside the wormhole may indeed provide for the possibility of sustenance of stable wormhole structure.
\end{abstract}

\maketitle

\section{Introduction}
\label{intro}
Wormholes at present are interstellar theoretical passages connecting two distant regions in spacetime, which have been intriguing the minds of physicists, since the advancement of general relativity (GR). These interesting structures are one of the many possible solutions to the Einstein field equations (EFE), which presents itself as a fascinating topic for systematic investigations in different technical frameworks.

A class of physically reasonable traversable wormhole structures are first described by Morris and Thorne in GR to explore the possibility of interstellar space travel without violating the laws of special relativity (SR)\cite{Morris1988}. Earlier studies demonstrated the possibility of traversable wormholes in GR \cite{Dehghani2009, Medizadeh2012}. A more technical definition of a traversable wormhole is a topological rupture in the fabric of spacetime that allows for traversable interstellar connections as well as other temporal links between distant spacetime points. A traversable wormhole with exotic matter leads one to expect that the energy conditions are violated in the wormhole. For instance, it was argued that Casimir energy can be a possible source to sustain a traversable wormhole geometry \cite{Garattini2019Nov,Jusufi2020}. Several works validating the requirement of violation of the energy conditions for the existence of wormholes can be found in \cite{Visser,Hochberg1997,Visser2003,Ida1999,Fewster2005,Kuhfitting2006, Zaslavskii2007,DeFalco2021Feb,DeFalco2021Feb1,DeFalco2020May,
DiGrezia2017Dec}. 
There are several important works, dealing with different frameworks, such as modified gravity and other physical coupling scenarios inspired from interaction with dark matter. Some of such work can be found in Refs. \cite{Chattopadhyay2023Dec,Capozziello2012Sep,
Errehymy2024Sep,Errehymy2024Jun,Errehymy2024Jun11,
Errehymy2024Aug,Errehymy2024Jun1,Errehymy2024May,
Errehymy2023Aug,Errehymy2023Aug1}.

Observational data suggests that a significant portion of the total matter content is probably occupied by dark energy, which is responsible for the accelerated expansion of the Universe. As a basis of this observational result, the background dynamics are provided by the GR (as well as modified theories of gravity). Detailed information can be found in references \cite{Ade2016,Delubac2015}. Generically, a cosmological fluid capable of generating negative pressure can be accounted as a possible explanation for the so-called dark energy. In the literature, there are several models of dark energy, which theoretically explain the accelerated expansion of the Universe with relevance to observed cosmological data \cite{Xia2016,Yang2017,Marsh2017, Sadjadi2006,Sadeghi2016,Khurshudyan2015a,Khurshudyan2015b,
Khurshudyan2016a,Khurshudyan2016b, Khurshudyan2017,Salucci2021Feb}. In the context of accelerated expansion of the Universe, \cite{Prigogine19} proposed a theory of  gravitational particle creation, which could be a competing candidate for explaining the observed accelerated expansion of the Universe and is based on non-equilibrium thermodynamics. This theory provides a mechanism capable of generating negative pressure in the Universe, through the production of particles involving thermodynamic processes at the expense of the gravitational field. Its primary advantage is that the negative pressure as a result of the creation of particles (also called \emph{back-reaction}) is naturally accommodated by the second law of thermodynamics. Since the particle creation process is macroscopically described by a negative back-reaction pressure of quantum origin, these
models have been studied as a possible means to explain the observed phenomena of the dark sector of cosmology \cite{Debnath2011,Steigman2009,Biswas2017}. Motivated by the aforementioned context, we suggest that the negative pressure produced as a result of the particle creation processes when incorporated into the energy-momentum tensor of the EFE, can be expected to contribute to the violation of the energy conditions required for the formation of wormholes. Therefore, we plan the paper in the following way: In section \ref{sec_review}, we briefly review the process of particle creation mechanism.  In section \ref{sec_deriv}, we derive the shape function of a wormhole in the presence of particle creation mechanism. In section \ref{embsec}, we obtained the 2D and 3D embedding visualization of the wormhole shape function obtained through the requirements of particle creation. In section \ref{sec_toyshape}, we derived the particle creation pressures inside the wormhole using various toy shape functions. In section \ref{conc}, we finally discuss and conclude the results of our study.

\section{Brief Review of Gravitational Particle Creation}
\label{sec_review}
Particle creation models have been thoroughly investigated as a promising framework for the purpose of describing the cosmological dark sector. This is achieved through matter creation process characterized by negative pressure of quantum origin, known as backreaction, which was mentioned in the previous section. Within this context, the accelerated expansion of the Universe in all eras, including the early time inflation as well as expansion of the Universe in the late-time era, emerges as a result of the mechanism of gravitational particle production. Some of the earlier development of this interesting possibility can be found in Refs. \cite{Lima1992a,Lima1992b,Lima2010Nov,Lima1996a,Lima1996,Freaza2002,Zimdahl2001,Silva2002Jun,
Jesus2014Jul,Mimoso2013Feb,Lima2012Nov,Jesus2011Sep,Basilakos2010Jul}. Cosmologically inspired wormholes are interesting and have been studied in various cosmological context \cite{Sengupta2023Sep,Pavlovic2023Mar,Kirillov2016May,Mehdizadeh2012Jun,
Sushkov2008Jan,Lemos2003Sep}. Quite recently, evolving wormholes in a Friedmann Universe have been studied by Bronnikov \cite{Bronnikov2023Oct}.  Motivated by this  likelihood of wormholes in various cosmological context, and relying on the assumption that a standard local wormhole spacetime can coexist locally in a global FLRW Universe, we intend to investigate the possibility of wormhole formation in a Friedmann Universe, within the background of gravitational particle creation mechanism.

To formulate the necessary equations of particle creation, one starts with the conservation equation of an open thermodynamic system of volume $V$ containing $N(t)$ number of particles, having energy density $\rho$ and thermodynamic pressure $p$. The conservation equation is 
\begin{equation}
d(\rho V) = dQ - pdV + \frac{h}{n}d(nV),
\label{term1}
\end{equation}
with $dQ$ as the heat recieved by the sytem in time $dt$. Here $h = \rho+ p$ is the enthalpy per unit volume and $n = N/V$ is the number density of particles. One may find that, in contrast to closed systems, where the particle number stays constant, the thermodynamic energy conservation in open systems incorporates a term (third term in Eq.\eqref{term1}) that represents the processes of matter creation and annihilation that occur within the system 
unlike isolated or closed systems where the number of particles is constant. 
The second law of thermodynamics takes the
following form 
\begin{equation}
dS = d_e S + d_i S \geq 0,
\label{pceq2}
\end{equation}
$d_e S$ denotes entropy flow, whereas $d_i S$ represents entropy generation. The first term of the equation reflects the system's homogeneity, while the latter relates to the entropy associated with particle creation. 
In order to obtain expressions for these aforementioned quantities, one may start by expressing the total differential of the entropy given by
\begin{equation}
\mathfrak{T} dS = d(\rho V) + pdV - \mu d(nV)
\label{pceq3}
\end{equation}
with $\mu$ as the chemical potential and $\mathfrak{T}$ being the thermodynamic temperature. Now, by utilizing Eq. \eqref{term1} and the standard relation $\mu n = h - \mathfrak{T}s$, where $s = S/V$ being the entropy density, it is possible to rewrite Eq. \eqref{pceq3} in the form 
\begin{equation}
\mathfrak{T}dS = dQ + \mathfrak{T}\frac{s}{n}d(nV).
\label{pceq4}
\end{equation}
From Eq. \eqref{pceq2}, we should have 
\begin{equation}
\mathfrak{T} dS = \mathfrak{T}d_e S + \mathfrak{T}d_i S,
\label{pceq5}
\end{equation}
Using Eq. \eqref{pceq5}, one can obtain direct relations for entropy flow and entropy creation as
\begin{equation}
d_s S = \frac{dQ}{\mathfrak{T}}, \quad d_i S = \frac{s}{n} d (nV).
\label{pceq6}
\end{equation}
Since the entropy flow may be considered as a measure of the variation in a system's homogeneity, a homogeneous system exhibits no variation in homogeneity. As an outcome, the entropy flow is zero in such a system. It suggests that in homogeneous systems, adiabatic processes are expected to occur, and hence matter formation is the primary source of entropy generation, i.e.
\begin{equation}
dS = d_i S = \frac{s}{n}d(nV) \geq 0.
\label{pceq7}
\end{equation}
This mechanism of irreversible creation of particles can be applied to cosmology. In a homogeneous and isotropic Universe, the volume of the Universe may be represented using the scale factor $a$ as $V = a^3(t)$.
Eq. \eqref{pceq1} may be recast in terms of total time derivatives of physical quantities as follows \cite{Prigogine19}:
\begin{equation} 
\frac{d}{dt^2} (\rho a^3) + p \frac{d}{dt}a^3 = \frac{dQ}{dt} + \frac{\rho + p}{n} \frac{d}{dt} (na^3).
\label{pceq8}
\end{equation}.
As homogenous systems do not receive heat, assuming the Universe is homogeneous, we can speculate that the heat received remains constant over time ($dQ/dt = 0$). This result enables us to reformulate Eq. \eqref{pceq8} in an equivalent form:
\begin{equation}
\dot{\rho} + 3H(\rho +p) = \frac{\rho + p}{n} (\dot{n} + 3Hn).
\label{pceq9}
\end{equation}
Eq. \eqref{pceq9} can be recast into a convenient form 
\begin{equation}
\dot{\rho} + 3H(\rho +p + \Pi) = 0,
\label{conser1}
\end{equation}
where \begin{equation}
\Pi = -\frac{\rho + p}{3Hn} (\dot{n} + 3Hn )
\label{pcp}
\end{equation}
is the particle creation pressure.
One may see that, the continuity Eq. \eqref{conser1} is a modified form of the usual continuity equation in GR, $\dot{\rho}+3H(\rho + p) =0$, where in an extra term $\Pi$ has been introduced as a consequence of matter creation. 
In an open thermodynamical system, the non-conservation of the total rate of change of the number of particles, $N = na^3$, with comoving volume $a^3$ and $n$ being the number density of particle production, yields \cite{Lima1992a, Lima1992b}
\begin{equation}
\dot{n} + 3Hn = n \Gamma,
\label{pceq10} 
\end{equation}
where $\Gamma$ is a particle creation rate (or source function) that can have negative or positive signs. Negative sign of $\Gamma$ represents the particle annihilation and positive sign of $\Gamma$ describes particle creation. Comparing Eqs. \eqref{pcp} and \eqref{pceq10}, the equation relating the creation rate $\Gamma$ to the creation pressure $\Pi$ can be found as 
\begin{equation}
\Pi := - \frac{\Gamma}{3H}(\rho + p).
\label{pcp2}
\end{equation}
In this work, we intend to investigate the existence of a traversable wormhole within a framework where the negative pressure required for sustaining the wormhole is provided by the particle creation mechanism given by \cite{Lima1992a,Lima1992b,Lima1996,Lima1996a,Zimdahl2001,Freaza2002}
\begin{equation}
\Pi := - \frac{\Gamma}{3H_0}(\rho + p) = -\frac{\Gamma \rho}{3H_0}(1+\omega),
\label{pc}
\end{equation}
where $\omega$ is the barotopic equation of state (EoS) parameter that relates the energy density with pressure, i.e. $p = \omega \rho$. Here, as we are investigating the wormhole solution in the present era, we have set $H = H_0$, which is the present value of the Hubble parameter. Moreover, we assume that the matter inside the wormhole is perfect fluid. In this work, we consider specifically that the perfect fluid follows a dust-like equation of state, i.e $\omega = 0$. 

\section{Wormhole Solution from Particle Creation Mechanism}
\label{sec_deriv}
Let us assume the spherically symmetric wormhole spacetime given by the metric
\begin{equation} 
d s^2=-e^{2 \Phi(r)} d t^2+\frac{d r^2}{1-b(r) / r}+r^2\left(d \theta^2+\sin ^2 \theta d \phi^2\right).
\label{metric}
\end{equation}
The metric functions $\Phi(r)$ and $b(r)$ are general functions of the radial coordinate $r$ within the context of a wormhole spacetime. The redshift function, denoted as $\Phi(r)$, characterizes the gravitational redshift effect, while the shape function, denoted as $b(r)$, provides information about the geometry of the wormhole.
It is essential to note that the radial coordinate $r$ exhibits non-monotonic behavior, wherein it decreases from infinity to a minimum value $r_0$, representing the location of the throat of the wormhole. At the throat, $b(r_0)=r_0$. Subsequently, the radial coordinate increases from $r_0$ to infinity.

Despite the presence of a coordinate singularity indicated by the divergence of the metric coefficient $g_{r r}$ at the throat, the proper radial distance $l(r)= \pm \int_{r_0}^r[1-b(r) / r]^{-1 / 2} d r$ must be finite across the entire wormhole geometry. This proper distance undergoes a continuous decrease from $l=+\infty$ in the upper universe to $l=0$ at the throat, and then further decreases from zero to $-\infty$ in the lower universe. For the case of a traversable wormhole, the required criteria are that the redshift function must be finite, i.e. $\exp{\Phi(r)} \neq 0$.

In this paper, we address wormhole formation within the context of irreversible matter creation formalism and thermodynamics of open systems in cosmology, as described by Prigogine \cite{Prigogine19}. This formalism views the Universe as an open system, with particle formation described by an adaptation of the energy-momentum tensor that incorporates a matter creation term into conservation laws. Consider an open system of volume $V$, containing $N(t)$ particles with energy density $\rho$ and thermodynamic pressure $p$.

In this work, we shall work with the commonly used form of the particle creation rate given by \cite{Lima1996}
\begin{equation}
\Gamma = 3\beta H_0,
\label{Gamm1}
\end{equation}
where $\beta$ is an arbitrary constant of the order of unity. According to the authors of \cite{Lima1996}, the values of $\beta << 1$ lead to negligible particle creation effects and may be neglected. The set of field equations can be obtained by using the metric \eqref{metric} in the Einstein field equations $G_{\mu \nu} = 8\pi T_{\mu \nu}$ as 
\begin{equation}
\begin{aligned}
\rho (r) &= \frac{b'}{8\pi r^2}, \\
p_r (r) &= - \frac{b}{r^3} - \Pi (r),\\
p_t(r) &= - \left(1 - \frac{b}{r}\right) \left(\frac{b'r - b}{2r^2(r-b)}\right) - \Pi(r).
\end{aligned}
\label{fes}
\end{equation}
where $\rho (r)$ is the energy density, $p_r(r)$ and $p_t(r)$ are the radial and tangential pressures respectively which are mutually orthogonal. $\Pi (r)$ is the particle creation pressure introduced into the energy-momentum tensor $T_{\mu \nu} = \text{diag}[- \rho(r), p_r (r) + \Pi(r), p_t (r) + \Pi(r), p_t (r) + \Pi(r)]$.
Subsituting $\rho$ and \eqref{Gamm1} into Eq. \eqref{pc} we obtain 
\begin{equation}
\Pi (r) = -\beta(1+\omega)\frac{b'(r)}{8\pi r^2},
\label{pc1}
\end{equation}
To obtain the explicit form of the shape function $b(r)$, the particle creation pressure should be specified in terms of the radial coordinate $r$. We argue that to sustain the wormhole structure the particle creation pressure needs to be larger at the throat (so that it can withstand the collapse of the throat) and then asymptotically decrease away from the throat. At the same time, the particle creation pressure should not be excessively large at the throat, as that would blow up the entire wormhole structure. Therefore, to obtain a sustained wormhole geometry, one must carefully chose a functional form of particle creation pressure $\Pi(r)$. For instance, if we assume the particle creation pressure in the form of inverse squared law like 
\begin{equation}
\Pi = - \frac{K}{r^2},
\label{pc2}
\end{equation}
where $K$ is a proportionality constant that accounts for the dimensions of $\Pi(r)$, one finds the shape function in a linear form
\begin{equation}
b(r) = r_0 + \frac{8\pi K}{\beta (1+\omega)} (r- r_0),
\label{shape}
\end{equation}
where we have substituted Eq. \eqref{pc2} in Eq. \eqref{pc1}.
Note that the linear shape function is a result of the inverse squared law given by Eq. \eqref{pc2}, which is the simplest case. To have a more generalized picture, we may chose the inverse power-law form, i.e.,
\begin{equation}
\Pi = - \frac{K}{r^\alpha},
\label{pc2_powlaw}
\end{equation}
where $\alpha > 0$ is a constant. The parameter $\alpha$, is tunable so that a non-linear form of shape function can be obtained. It is also seen that for Thus solving Eq. \eqref{pc1} with the help of Eq. \eqref{pc2_powlaw}, we get the general non-linear shape function 
\begin{equation}
b(r) = r_0 -\frac{8 \pi  K \left(r^{3-\alpha }-r_0^{3-\alpha }\right)}{\beta  (\alpha -3) (\omega +1)}; \quad \alpha \neq 3, \omega \neq -1.
\label{non-lin_shape}
\end{equation}
The shape function given by Eq. \eqref{non-lin_shape} determines the configuration of the wormhole which is sustained by the negative particle creation pressure. The constraint on the possible values of $\alpha$ are $(0,3) \cup (3,\infty)$. However, to keep things simple, we choose to confine ourselves to the range $(0,3)$. In this work, we are motivated to investigate the possibility of a sustained wormhole structure in the presence of particle creation mechanism, without an external introduction of exotic fluids to account for the negative pressure required to sustain the wormhole geometry.

From Eq. \eqref{fes}, given the shape function \eqref{non-lin_shape}, one may obtain the energy density, radial and tangential pressures given by 
\begin{equation}
\rho (r) = \frac{K r^{-\alpha }}{\beta  \omega +\beta },
\label{rho_mod1}
\end{equation}
\begin{equation}
p_r (r) = \frac{8 \pi  K r^{-\alpha }}{(\alpha -3) \beta  (\omega +1)}+K r^{-\alpha }-\frac{8 \pi  K r_0^{3-\alpha }}{(\alpha -3) \beta  r^3 (\omega +1)}-\frac{r_0}{r^3},
\label{rad_tens1}
\end{equation}
\begin{equation}
p_t(r) = \frac{r_0^{-\alpha } r^{-\alpha -3} \left(8 \pi  K r_0^3 r^{\alpha }-2 K r^3 r_0^{\alpha } (4 \pi  (\alpha -2)-(\alpha -3) \beta  (\omega +1))\right)}{2 \beta (\alpha -3)  (\omega +1)}+\frac{r_0}{2 r^3},
\label{tang_press_1}
\end{equation}
\subsection{Constraints on $\beta$: Stability analysis}
\begin{figure*}[tbh]
\centerline{
\includegraphics[scale=0.45]{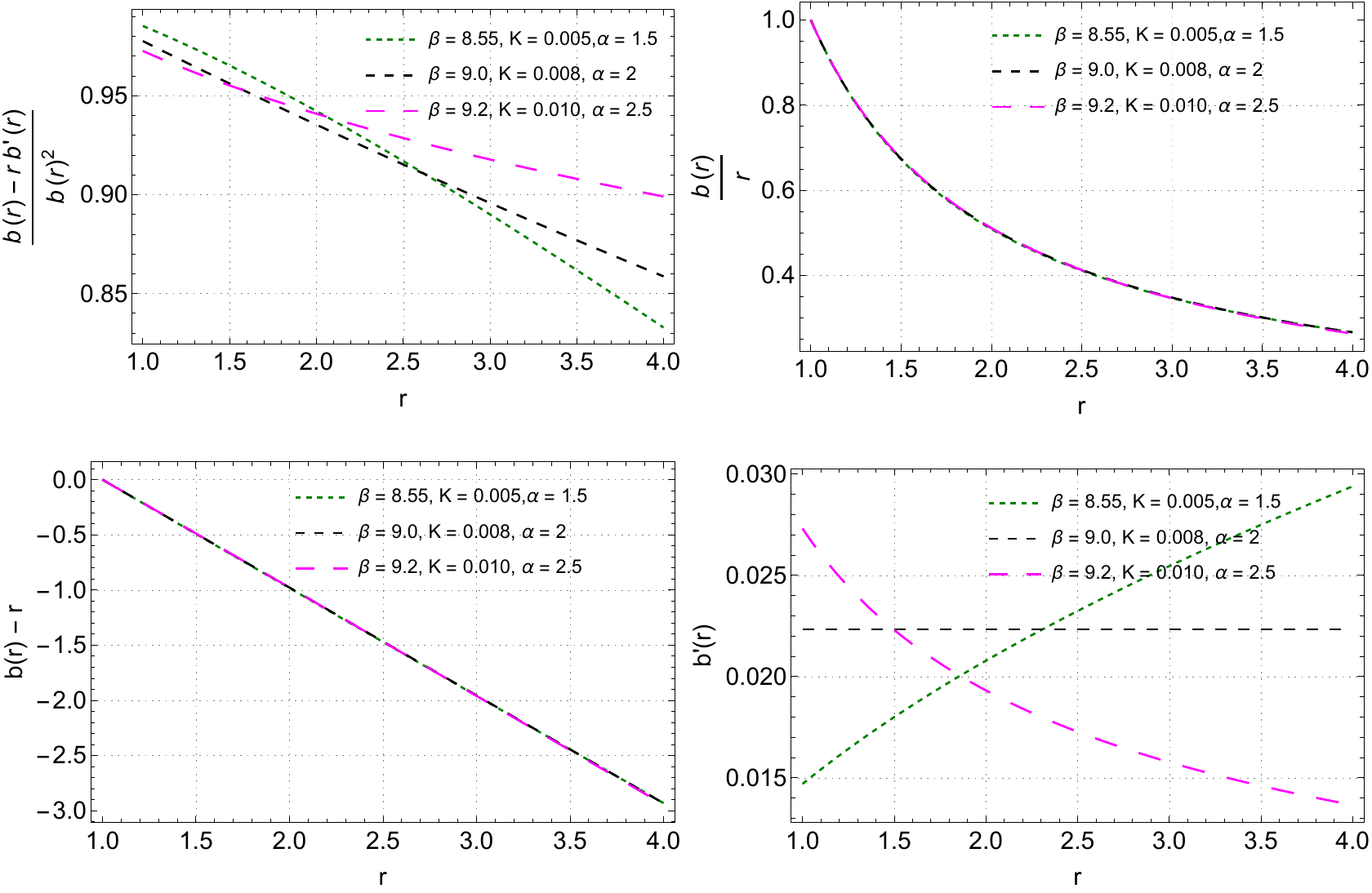}}
\caption{The viability conditions related to the shape function \eqref{non-lin_shape} is shown for three values of $K$, $\alpha$ and $\beta$}.
\label{viab_con1}
\end{figure*}

It is necessary to choose the parameter $\beta$ associated with particle creation in such a way that the obtained wormhole structure exhibits stability. In order to confirm the stability of the wormhole, we adopt the method of squared adiabatic sound speed given by \cite{Battista2024Sep}
\begin{equation}
v_s^2 = \frac{d\langle p \rangle}{dr}\left(\frac{d\rho}{dr}\right)^{-1}.
\label{sound}
\end{equation}
where $\langle p \rangle = \frac{1}{3}(p_r + 2p_t)$, is the averaged pressure taken across the three spatial dimensions. This physical quantity represents the speed at which tiny acoustic perturbations or sound waves propagate through a particular medium during an adiabatic process. This allows us to probe into the stability of the wormhole in response to these small perturbations. The causality condition dictates $0 \le v_s^2 < 1$, which means that the speed limit of light cannot be exceeded by the perturbations. Using Eqs. \eqref{rho_mod1}, \eqref{rad_tens1} and \eqref{tang_press_1} we obtain 
\begin{equation}
v_s^2 = \beta -\frac{8 \pi }{3}
\label{vs_mod1}
\end{equation}
To respect the causality condition we must have the following constraint on $\beta$
\begin{equation}
\frac{8 \pi }{3}\leq \beta <1+\frac{8 \pi }{3}, \quad \text{or} \quad 8.373 \leq \beta < 9.373.
\label{constraint1}
\end{equation}
Having obtained the constrained values of $\beta$ within the range \eqref{constraint1}, we can now choose the values for $\beta$, that would ensure the stability of the wormhole. Let us now examine the validity of our derived shape function \eqref{non-lin_shape}, through the viability conditions. The necessary conditions for a viable wormhole geometry are 
\begin{enumerate}
\item \emph{Throat condition:} At the throat $r_0$, $b\left(r_0\right)=r_0$ and $ 1 - \frac{b(r)}{r} > 0$ or $b(r) - r < 0$ outside the throat.
\item \emph{Flaring out condition:} $b^{\prime}\left(r_0\right)<1$ i.e. \[\frac{b(r)-r b^{\prime}(r)}{b^2(r)}> 0,\] where (') represents derivative w.r.t. $r$.
\item \emph{Asymptotically Flatness condition:} \[\frac{b(r)}{r} \rightarrow 0 \]as $r \rightarrow \infty.$
\end{enumerate}
Fig. \ref{viab_con1} shows the throat, flareout and asymptotic conditions are fulfilled, with three sets of the parameter values $\beta$ and $K$. The quantities $b'(r)$, $b(r)/r$, $b(r)-r$ and $\frac{b(r)-r b^{\prime}(r)}{b^2(r)}$ are plotted for different combination of $\beta$, $K$ and $\alpha$. We can see clearly from the plots that the shape function \eqref{non-lin_shape} describes a viable wormhole geometry by adhering to the aforementioned conditions.

To explore the physical nature of the matter content inside the wormhole geometry, we shall make use of the energy conditions. A generic aspect of wormhole systems is the violation of the null energy condition (NEC) (i.e. $\rho + p_r \ge 0$ or $\rho + p_t \ge 0$ in terms of radial and tangential pressures respectively) at the throat of the wormhole structure. On the other hand, for some specific models keeping valid the NEC while keeping the dominant energy condition (DEC) ($\rho - p_r  \ge 0$ or $\rho - p_t \ge 0$) violated at the throat is also possible \cite{Rosa2018,Ayan,Tangphati,Lu2024,Lobo2006Mar}. In this work, we shall construct traversable wormhole structures with the violation of the NEC as well as DEC at the throat of the wormhole, simultaneously keeping the energy condition $\rho \ge 0$. This turns out, as a result of the violation of the weak energy condition (WEC) ($\rho \ge 0, \rho + p_r \ge 0 \text{ or } \rho + p_t \ge 0$).

To study the energy conditions for the wormhole geometry \eqref{non-lin_shape}, let us calculate the quantities $\rho + p_r$, $\rho + p_t$, $\rho - p_r$ and $\rho - p_t$ given by
\begin{equation}
\rho + p_r := \frac{K (\beta  \omega +\beta +1) r^{-\alpha }}{\beta  (\omega +1)}+\frac{8 \pi  K \left(r^{-\alpha }-\frac{r_0^{3-\alpha }}{r^3}\right)}{\beta (\alpha -3)  (\omega +1)}-\frac{r_0}{r^3},
\label{nec1}
\end{equation}
\begin{equation}
\rho + p_t :=\frac{r^{-\alpha -3} r_0^{-\alpha } \left(8 \pi  K r_0^3 r^{\alpha }+r_0^{\alpha } \left(2 K r^3 ((\alpha -3) (\beta  \omega +\beta +1)-4 \pi  (\alpha -2))+(\alpha -3) \beta  r_0 (\omega +1) r^{\alpha }\right)\right)}{2\beta (\alpha -3)   (\omega +1)},
\label{nec11}
\end{equation}
and 
\begin{equation}
\rho - p_r := -\frac{K (\beta  \omega +\beta -1) r^{-\alpha }}{\beta  (\omega +1)}-\frac{8 \pi  K \left(r^{-\alpha }-\frac{r_0^{3-\alpha }}{r^3}\right)}{\beta(\alpha -3) (\omega +1)}+\frac{r_0}{r^3}.
\label{dec1}
\end{equation}
\begin{equation}
\rho - p_t := \frac{r^{-\alpha -3} r_0^{-\alpha } \left(r_0^{\alpha } \left(2 K r^3 (4 \pi  (\alpha -2)-(\alpha -3) (\beta  \omega +\beta -1))-(\alpha -3) \beta  r_0 (\omega +1) r^{\alpha }\right)-8 \pi  K r_0^3 r^{\alpha }\right)}{2 \beta  (\alpha -3) (\omega +1)}.
\label{dec11}
\end{equation}


\begin{figure*}[tbh]
\centerline{
\includegraphics[scale=0.45]{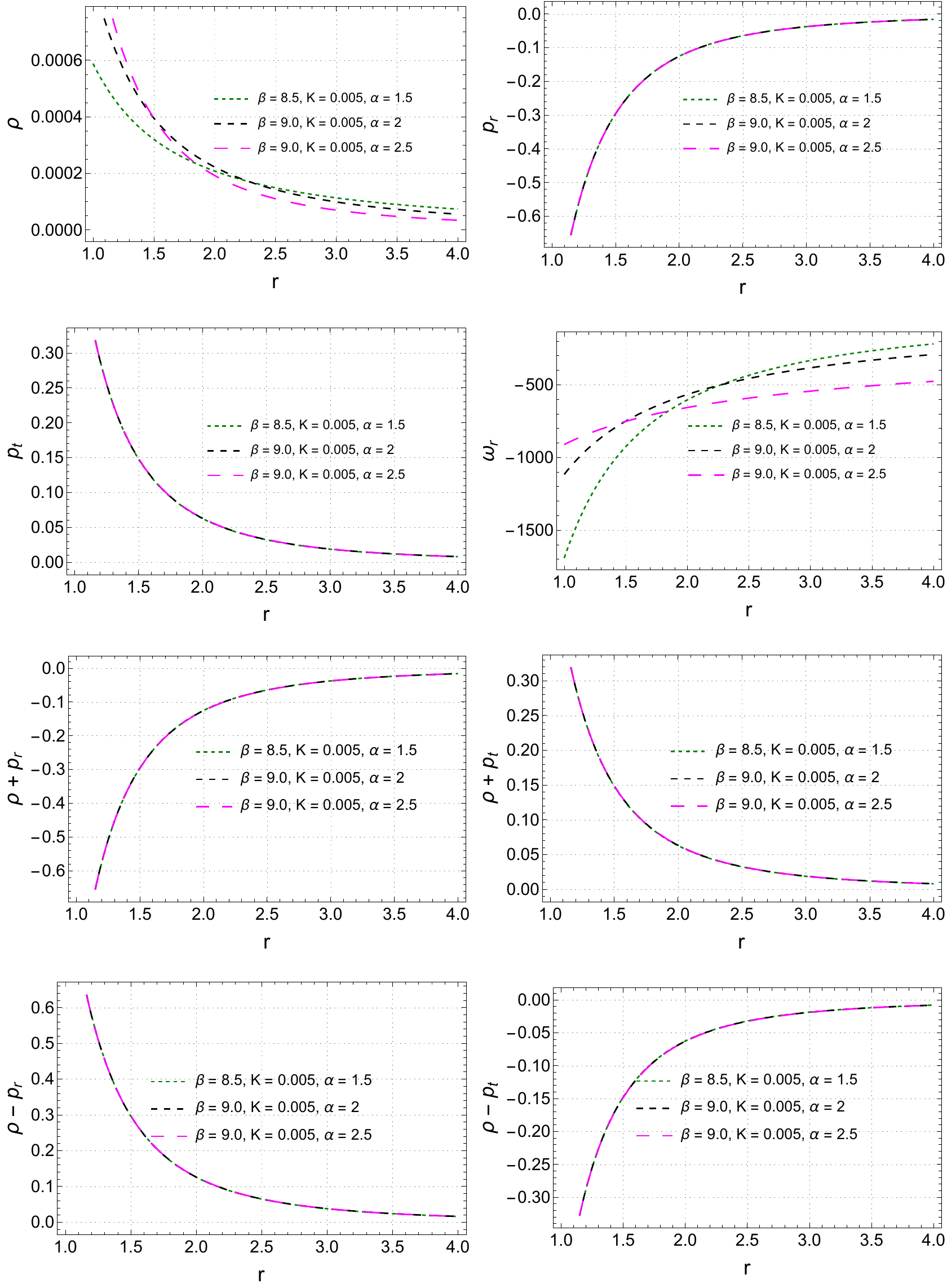}}
\caption{The physical parameters and the energy conditions for the shape function \eqref{non-lin_shape} is shown for three combinations of $K$, $\alpha$ and $\beta$}.
\label{en_cond_shape1}
\end{figure*}
Fig. \ref{en_cond_shape1} displays the behaviour of the related physical parameters $\rho(r), p_r (r), p_t(r) \text{ and } \omega_r = p_r(r)/\rho$ and the energy conditions like NEC and DEC for different combinations of the model parameters. As mentioned before, for the analysis we have set the EoS of the matter content inside the wormhole to follow a dust like $\omega = 0$.  Moreover, we shall choose the throat radius $r_0 = 1$. This choice is arbitrarily positive, provided the energy viability conditions are satisfied \cite{Jusufi2020Aug,Mustafa2024Jan}. The energy density as expected always takes positive values and is maximum at the throat and decreases asymptotically away from the throat. The radial pressure always takes negative values, which may be attributed to counteracting the radial collapse of the wormhole structure. Also, the tangential pressure always remain positive. The NEC in terms of radial pressure shows clear violation at the throat but obeys at far away regions from the throat, whereas the DEC is obeyed at the throat as well as at far away regions. However, in terms of the tangential pressure, the NEC is obeyed at the throat as well as in asymptotic regions but DEC is violated at the throat and satisfied at the asymptotic regions. Also as $\rho \ge 0$, one can say WEC is violated for the radial pressure case and obeyed for the tangential pressure one.  To examine the behaviour of the effective matter resulting from the particle creation mechanism, we calculated the radial EoS parameter $\omega_r = p_r /\rho$, that essentially evolves with negative values while away from the throat it increases. This indicates that the effective matter at the throat behaves like exotic matter. In other words, we find that without explicitly introducing non-exotic matter, the wormhole is sustained due to the presence of particle creation mechanism in the interior of the wormhole structure. The negative pressure that arises at the throat is entirely due to particle creation mechanism.

\section{Embedding}
\label{embsec}
\begin{figure*}[!tpbh]
\centerline{\includegraphics[scale=0.45]{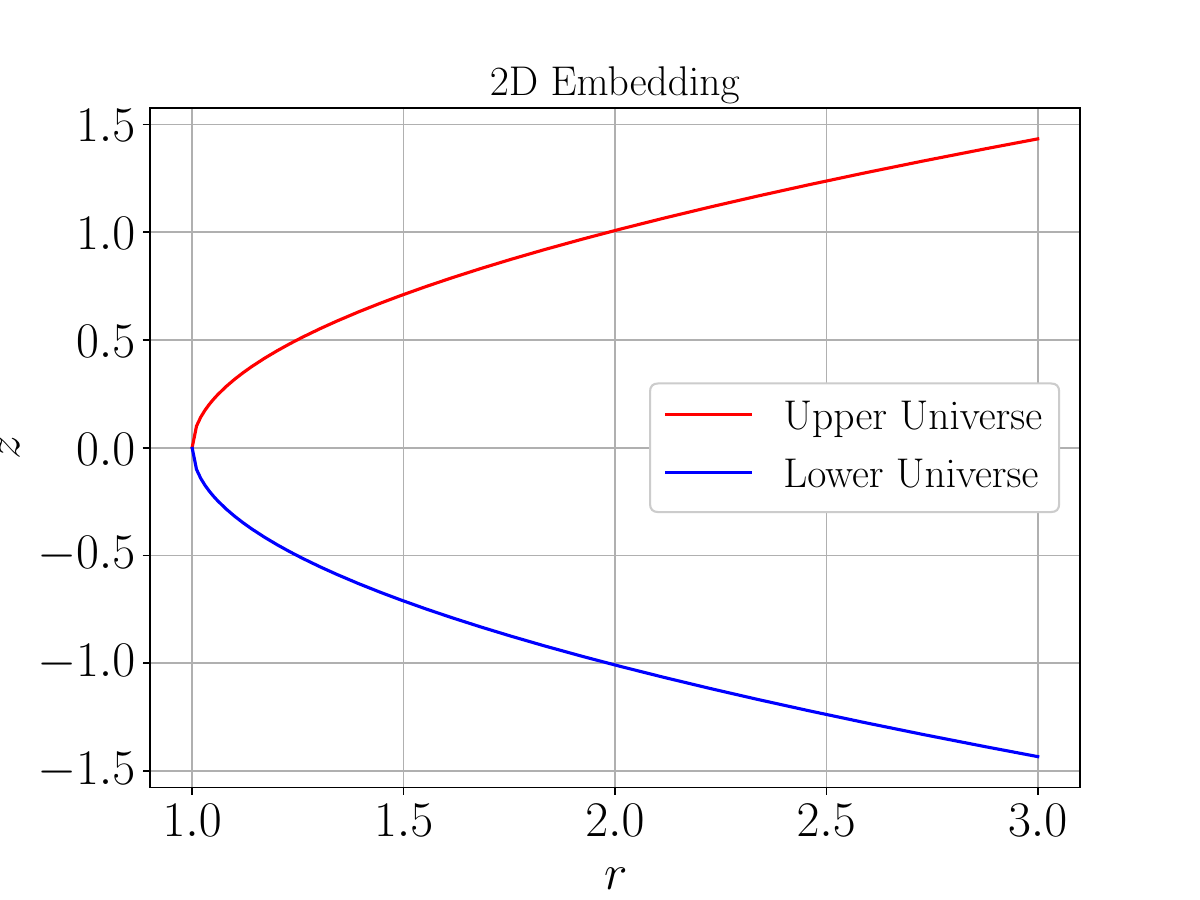} \hspace{-2.5cm}
\includegraphics[scale=0.45]{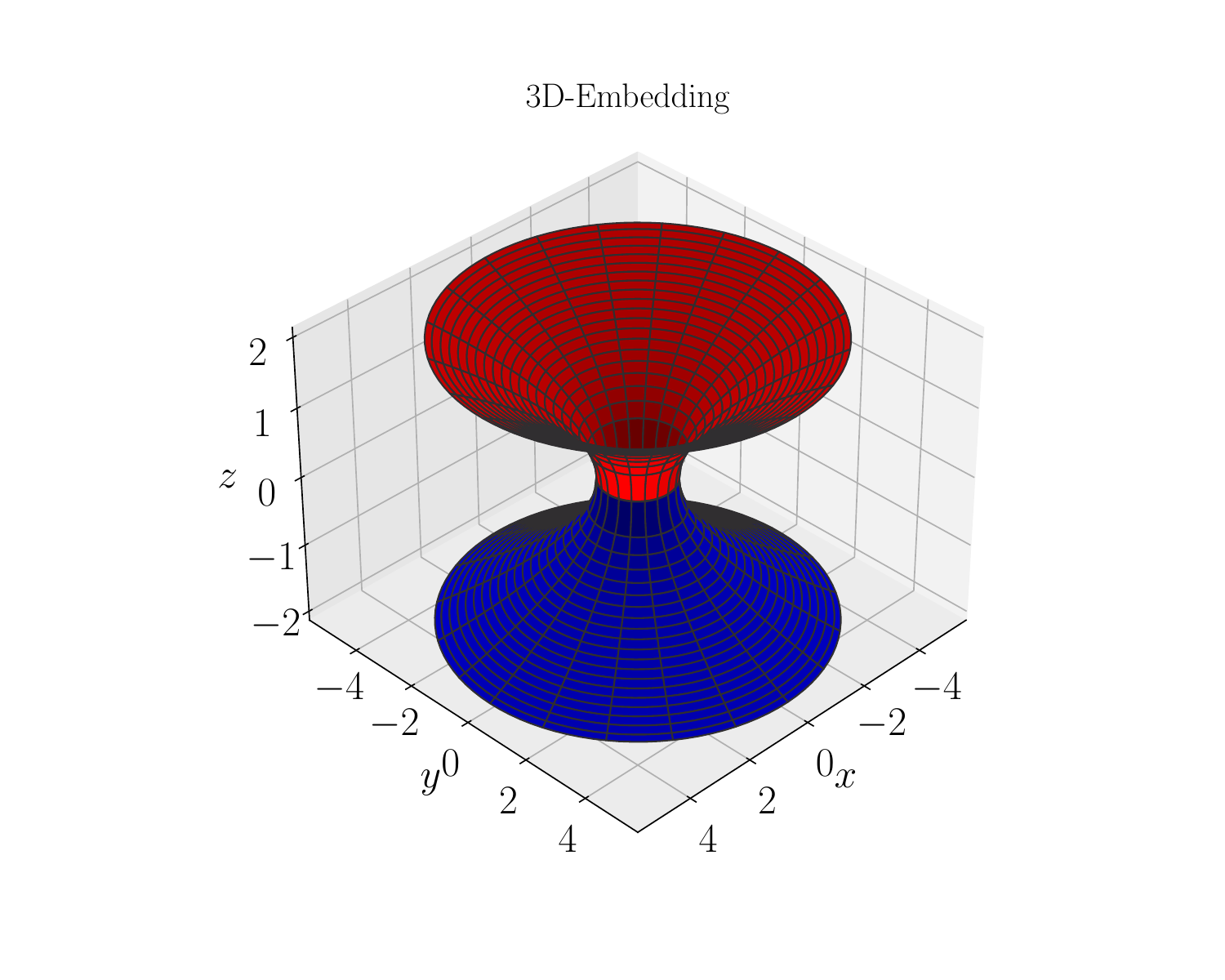}}
\caption{The 2D and 3D embedding of the wormholes are shown for the shape functions \eqref{shape} the throat is assumed at $r_0 = 1$ and the values of $\beta$ and $K$ is taken to be $8.55$ and $0.005$ respectively.}
\label{embed}
\end{figure*}

To realize the embedding function of a wormhole, one may choose the equatorial slice $\theta = \pi/2$ and $t = $constant. This reduces the metric \eqref{metric} to the form
\begin{equation}
ds^2 = \left(1 - \frac{b(r)}{r}\right)^{-1} dr^2 + r^2 d\phi^2.
\label{red_metric}
\end{equation}
The metric \eqref{red_metric} is equivalent to 
\begin{equation}
ds^2 = dz^2 + dr^2 + r^2 d\phi^2,
\label{metr_cyl}
\end{equation}
in cylindrical coordinates. The embedded surface may be related to a 3D-Euclidean surface defined by $z = z(r)$, the metric \eqref{metr_cyl} can be further expressed in the following form
\begin{equation}
ds^2 = \left[ 1 + \left( \frac{dz}{dr} \right)^2 \right] dr^2 + r^2 d\phi^2,
\label{3d_euc} 
\end{equation}
which can now be compared with Eq. \eqref{red_metric} to obtain 
\begin{equation}
\frac{dz}{dr} = \pm \left(\frac{r}{b(r)} - 1\right)^{-1/2},
\label{dzdr}
\end{equation}
Eq. \eqref{dzdr} can be solved to obtain a functional form of $z(r)$. However in our model given Eq. \eqref{non-lin_shape}, the analytical form of a general $z(r)$ is not obtainable because of the complicated form of the shape function. Therefore, we adopt a numerical method to visualize the embedding function, which is shown in Fig. \ref{embed}.

The embedding function obtained for the shape function \eqref{non-lin_shape}, obtained from physical consideration of particle creation mechanism is shown in the Fig. \ref{embed} for $\beta = 8.55$ and $K = 0.005$.
The embedding function and the diagrams for the toy models \eqref{shape2},\eqref{shape3} and \eqref{shape4} can be found in \cite{Ghosh,Samanta2020,Godani2019Jul}. 

\section{Toy Shape Functions and Evolution of Particle Creation Pressure}
\label{sec_toyshape}
In the previous discussion, we demonstrated how the presence of particle creation mechanism at the throat provides for a possibility of a stable wormhole geometry. We utilized a inverse powerlaw variation of the particle creation pressure along the wormhole structure and derived a general shape function describing the geometry of the wormhole. Now, in this section, we are going to study the reverse process where we assume some popularly used shape functions and investigate how the particle creation pressure should vary within the wormhole. There have been several works on possibility of wormholes in GR through the use of wormhole shape functions \cite{Rahaman2019Jan, Konoplya2022Mar,Armendariz-Picon2002Apr}. Specifically we deal with three shape functions \textbf{(i)} $b(r) = \frac{r_0 \tanh(\epsilon r)}{\tanh(\epsilon r_0)}$, \textbf{(ii)}  $b(r) =r_0^n r^{1-n},$ and \textbf{(iii)} $b(r) = r \exp [\Omega(r - r_0)]$. The detailed analysis is carried out below:
\subsection{Shape function 1: $b(r) = \frac{r_0 \tanh(\epsilon r)}{\tanh(\epsilon r_0)}$}
Let us now assume a hyperbolic shape function given by
\begin{equation}
b(r) = \frac{r_0 \tanh(\epsilon r)}{\tanh(\epsilon r_0)},
\label{shape2}
\end{equation}
where $\epsilon$ is a constant parameter. 
This form of shape function has been studied in different contexts like Lyra manifold \cite{Moradpur} and $f(R)$ gravity \cite{Ghosh}.

\begin{figure*}[tbh]
\centerline{
\includegraphics[scale=0.45]{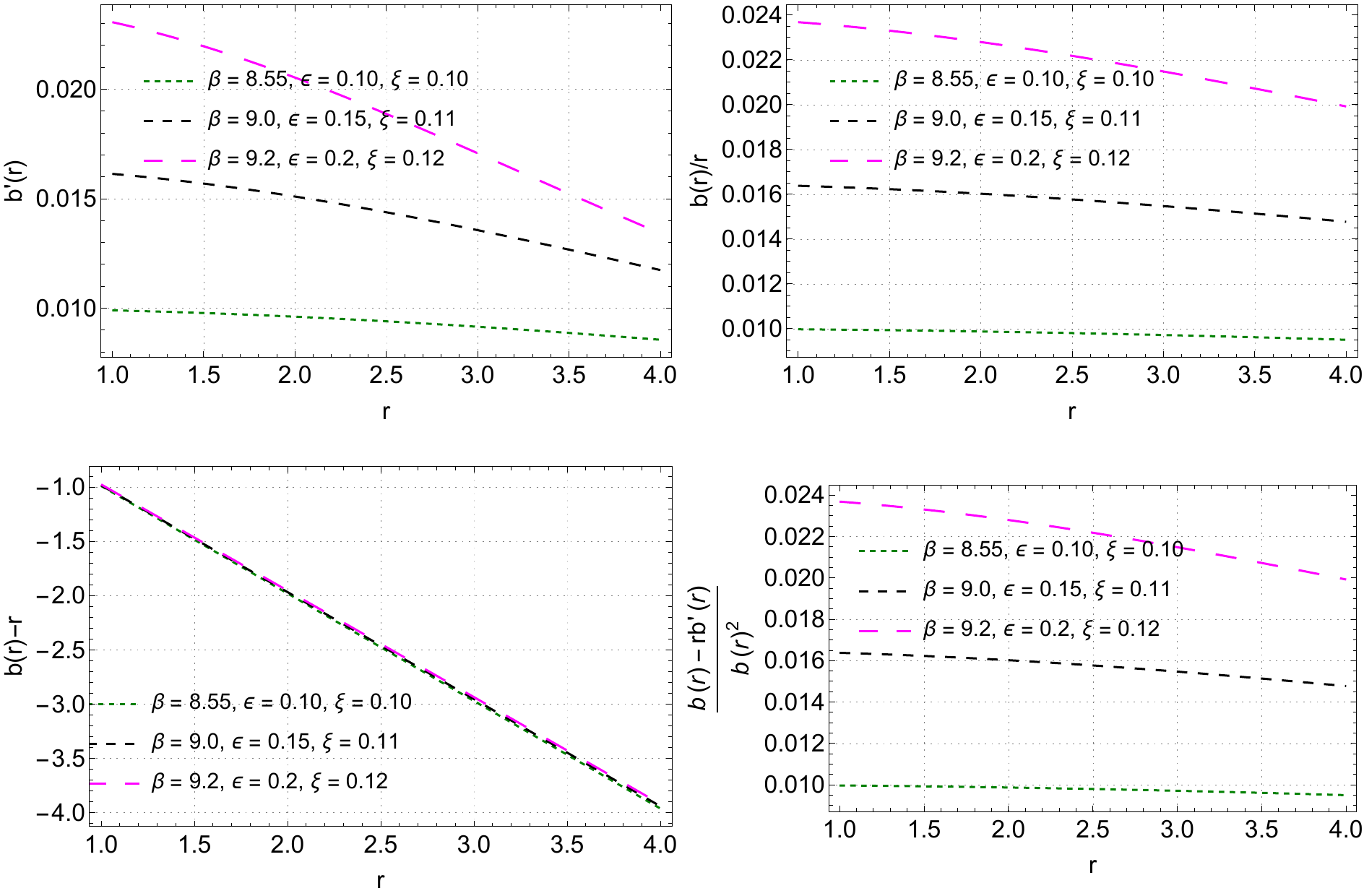}}
\caption{The viability conditions related to the shape function \eqref{shape2} is shown for three values of $\epsilon$ and $\beta$.}
\label{cond_sh1}
\end{figure*}

The energy density with this shape function is
\begin{equation}
\rho = \frac{r_0 \epsilon  \sech^2(r \epsilon ) \coth (r_0 \epsilon )}{8 \pi  r^2},
\label{rho2}
\end{equation}
which clearly is a positive function for all values of $r$.
In Fig.\ref{cond_sh1}, $b'(r)$, $b(r)/r$, $b(r)-r$ and $\frac{b(r)-r b^{\prime}(r)}{b^2(r)}$ are plotted for different values of $\epsilon$, $\xi$ and $\beta$. As required by a viable wormhole structure, the conditions mentioned in section \ref{sec_deriv} are clearly obeyed by the assumed shape function.
The radial pressure $p_r$ and tangential pressure $p_t$ in this case take the form
\begin{equation}
 p_r  = \frac{r_0 \sech^2(r \epsilon ) \coth (r_0 \epsilon ) (\beta  r (\omega +1) \epsilon -4 \pi  \sinh (2 r \epsilon ))}{8 \pi  r^3},
 \label{p2}
\end{equation}
\begin{equation}
p_t = \frac{r_0 \sech^2(r \epsilon ) \coth (r_0 \epsilon ) (r \epsilon  (\beta  \omega +\beta -4 \pi )+2 \pi  \sinh (2 r \epsilon ))}{8 \pi  r^3}.
\label{p2t}
\end{equation}
The particle creation pressure for the shape function \eqref{shape2} can be obtained from Eq. \eqref{pc1}
\begin{equation}
\Pi (r) = -\frac{\beta r_0 (\omega +1) \epsilon  \sech^2(r \epsilon ) \coth (r_0 \epsilon )}{8 \pi  r^2}
\label{Pi_2}
\end{equation}
Before proceeding further, let us first determine the constraints on the allowed values of $\beta$, through the analysis of squared sound speed. The sound speed is found to be 
\begin{equation}
v_s^2 = \beta -\frac{8 \pi }{3}
\label{vs_mod2}
\end{equation}
It is found that the expression for sound speed remains unchanged, given by Eq. \eqref{vs_mod1} irrespective of the shape function (and hence $\rho$, $p_r$ and $p_t$). Thus we see that the sound speed is strictly dictated by the parameter $\beta$ only. This results in the same constraints on $\beta$ given by Eq. \eqref{constraint1}. Thus, we may use the same values of $\beta$ just like the previous case, in the analysis concerned with the toy shape functions.

\begin{figure}[!tbh]
\centerline
\centerline{\includegraphics[scale=0.5]{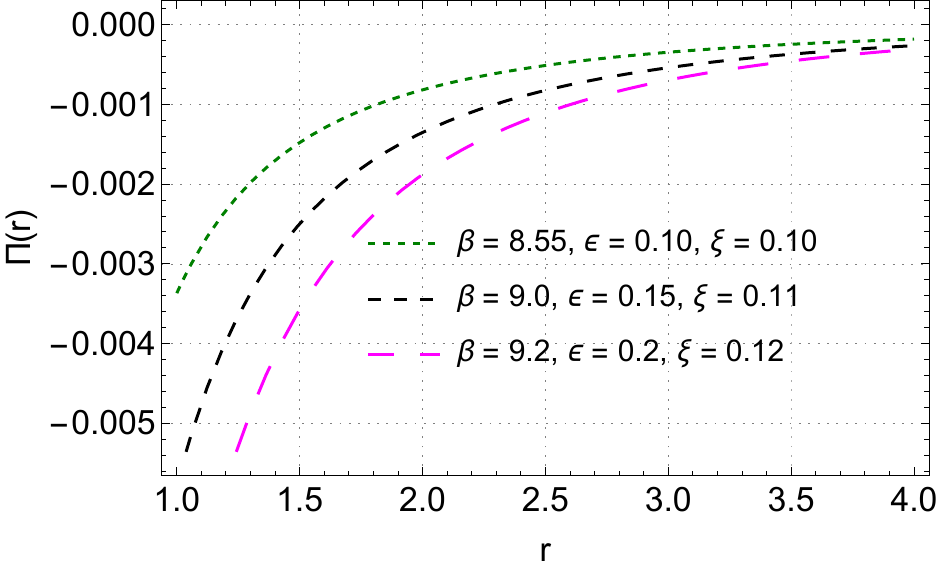}}
\caption{Particle creation pressures with shape function \eqref{shape2}}.
\label{fig_pc_1}
\end{figure}

Now, the quantities related to the energy conditions are found to be
\begin{widetext}
\begin{equation}
\rho + p_r := \frac{r_0 \sech^2(r \epsilon ) \coth (r_0 \epsilon ) (r \epsilon  (\beta  \omega +\beta +1)-4 \pi  \sinh (2 r \epsilon ))}{8 \pi  r^3},
\label{nec2}
\end{equation}
\begin{equation}
\rho + p_t := \frac{r_0 \sech^2(r \epsilon ) \coth (r_0 \epsilon ) (r \epsilon  (\beta  \omega +\beta -4 \pi +1)+2 \pi  \sinh (2 r \epsilon ))}{8 \pi  r^3},
\label{nec22}
\end{equation}
\begin{equation}
\rho - p_r := \frac{r_0 \sech^2(r \epsilon ) \coth (r_0 \epsilon ) (4 \pi  \sinh (2 r \epsilon )-r \epsilon  (\beta  \omega +\beta -1))}{8 \pi  r^3},
\label{dec2}
\end{equation}
\begin{equation}
\rho - p_t := -\frac{r_0 \sech^2(r \epsilon ) \coth (r_0 \epsilon ) (r \epsilon  (\beta  \omega +\beta -4 \pi -1)+2 \pi  \sinh (2 r \epsilon ))}{8 \pi  r^3},
\label{dect2}
\end{equation}
\end{widetext}
\begin{figure*}[!th]
\centerline{\includegraphics[scale=0.45]{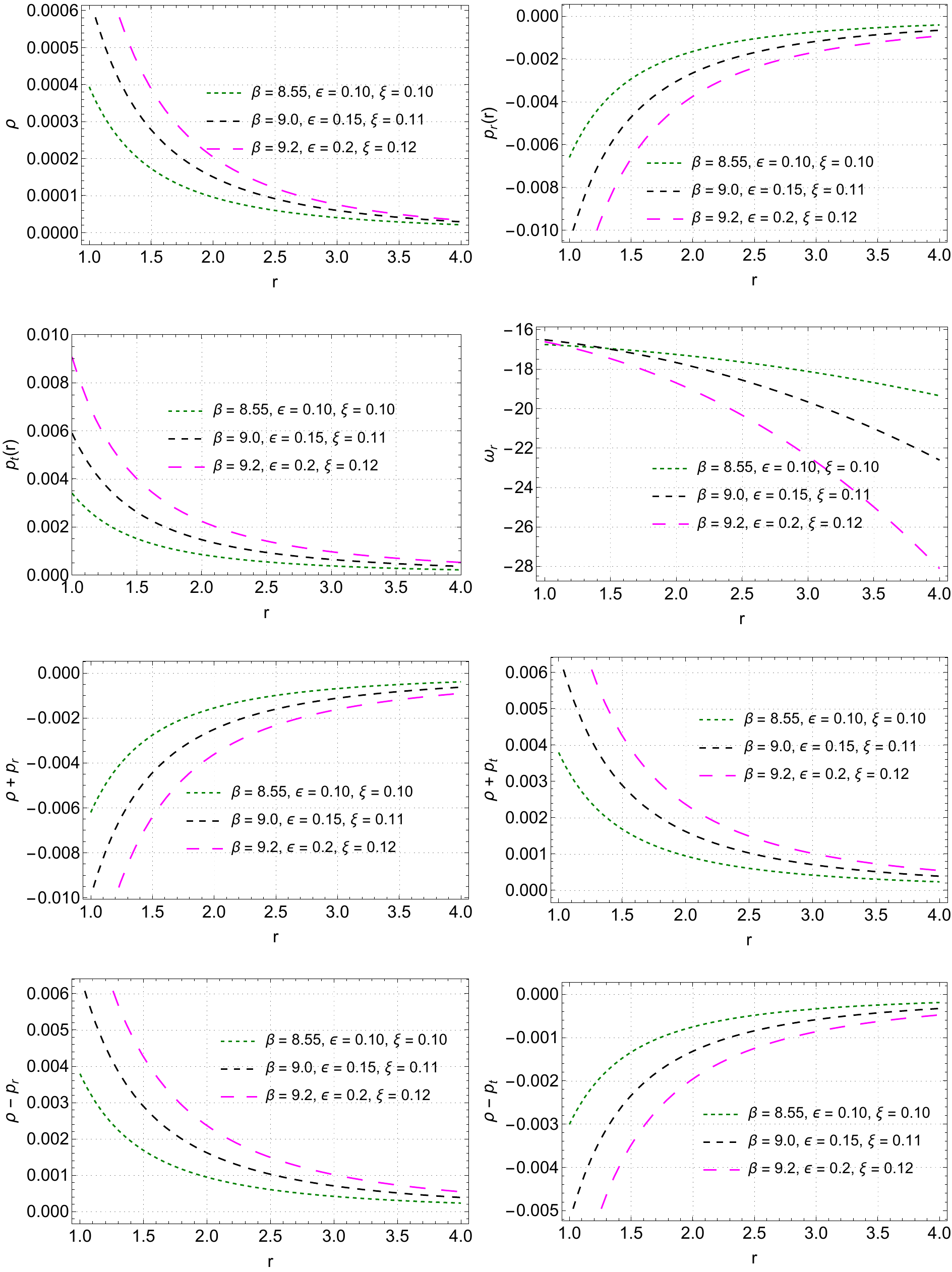}} \caption{The physical parameters and the energy conditions for the shape function \eqref{shape2} is shown for three values of $\epsilon$ and $\beta$.}
\label{fig_ec_2}
\end{figure*}
From Fig. \ref{fig_ec_2}, it is observed that the energy density $\rho$ for this model in the presence of particle creation decreases radially, with maximum value at the throat. 
The evolution of the radial pressure $p_r$ is observed, with maximum negative values at the throat with asymptotically vanishing at far regions. This negative pressure may be thought of as a means of preventing the radial collapse of the wormhole at the throat. However, the tangential pressure remains positive. NEC is violated in terms of radial pressure at the throat but tends to obey at the asymptotic limit, whereas the DEC is obeyed at the throat as well as at the asymptotic limit. In terms of the tangential pressure, the NEC is obeyed at the throat as well as at asymptotic regions but DEC is obeyed at the throat. Also, as $\rho \ge 0$, WEC is violated for the radial pressure case and obeyed for the tangential pressure one. Moreover, from Eq. \eqref{Pi_2} and Fig. \ref{fig_pc_1}, it is evident that the particle creation pressure is negative at the throat all the way to $r \to \infty$, which is required and expected for backreaction pressure responsible for sustaining a wormhole geometry. To further examine the behaviour of the effective matter resulting from the particle creation mechanism, the radial EoS parameter $\omega_r = p_r /\rho$, is also plotted that evolves with negative values, that shows exotic matter like behaviour at the throat while taking higher negative values away from the throat.
\subsection{Shape function 2: $b(r) = r_0^n r^{1-n}$}
\begin{figure*}[tbh]
\centerline{
\includegraphics[scale=0.45]{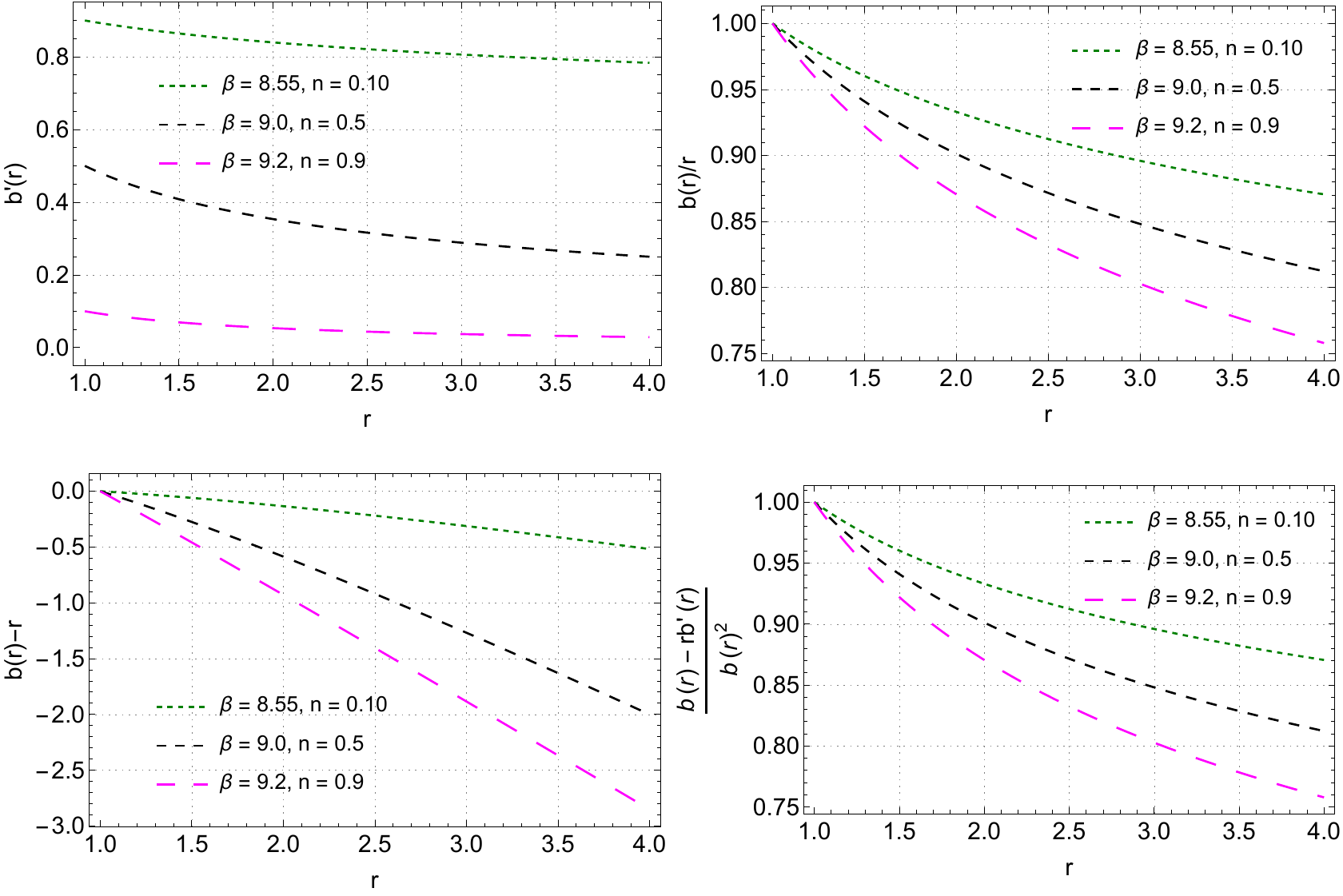}}
\caption{The viability conditions related to the shape function \eqref{shape3} is shown for three values of $n$ and $\beta$.}
\label{cond_sh2}
\end{figure*}

Another viable wormhole is a power-law type shape function given by
\begin{equation}
b(r) =r_0^n r^{1-n},
\label{shape3}
\end{equation}
where $n$ is a model parameter and $r_0$ is the throat radius. 
This type of shape function has been studied in \cite{Ghosh}, in the context of $f(R)$ gravity reconstruction for given choices of wormhole shape functions.
In Fig. \ref{cond_sh2} $b(r)$, $b(r)/r$, $b(r)-r$ and $\frac{b(r)-r b^{\prime}(r)}{b^2(r)}$ are plotted for different values of $n$. It is seen that the required criteria for validity of a wormhole shape function are satisfied for different combinations of $n$.
The energy density for this particular shape function takes the form
\begin{equation}
\rho = \frac{(1-n) r^{-n-2} r_0^n}{8 \pi },
\label{rho3}
\end{equation}
The radial and tangential pressure can be calculated as
\begin{equation}
 p_r (r) = \frac{\beta  (1-n) (\omega +1) r^{-n-2} r_0^n}{8 \pi }-r^{-n-2} r_0^n,
 \label{p3}
\end{equation}
\begin{equation}
 p_t(r) = \frac{r^{-n-2} r_0^n (\beta  (\omega +1)-\beta  n (\omega +1)+4 \pi  n)}{8 \pi },
 \label{pt3}
\end{equation}
The particle creation pressure can be obtained as
\begin{equation}
\Pi (r) = -\frac{\beta  (1-n) (\omega +1) r^{-n-2} r_0^n}{8 \pi}.
\label{Pi_3}
\end{equation}
Also, we have the quantities related to the energy conditions
\begin{equation}
\rho + p_r := \frac{r^{-n-2} r_0^n (\beta  \omega +\beta -n (\beta  \omega +\beta +1)-8 \pi +1)}{8 \pi },
\label{nec3}
\end{equation}
\begin{equation}
\rho + p_t := \frac{r^{-n-2} r_0^n (\beta  \omega +\beta +n (-\beta  (\omega +1)+4 \pi -1)+1)}{8 \pi },
\label{nect3}
\end{equation}
\begin{equation}
\rho - p_r:= \frac{r^{-n-2} r_0^n (\beta  (-\omega )-\beta +n (\beta  \omega +\beta -1)+8 \pi +1)}{8 \pi },
\label{dec3}
\end{equation}
\begin{equation}
\rho - p_t := -\frac{r^{-n-2} r_0^n (\beta  \omega +\beta +n (\beta  (-\omega )-\beta +4 \pi +1)-1)}{8 \pi },
\label{dect3}
\end{equation}
\begin{figure*}[!th]
\centerline{\includegraphics[scale=0.45]{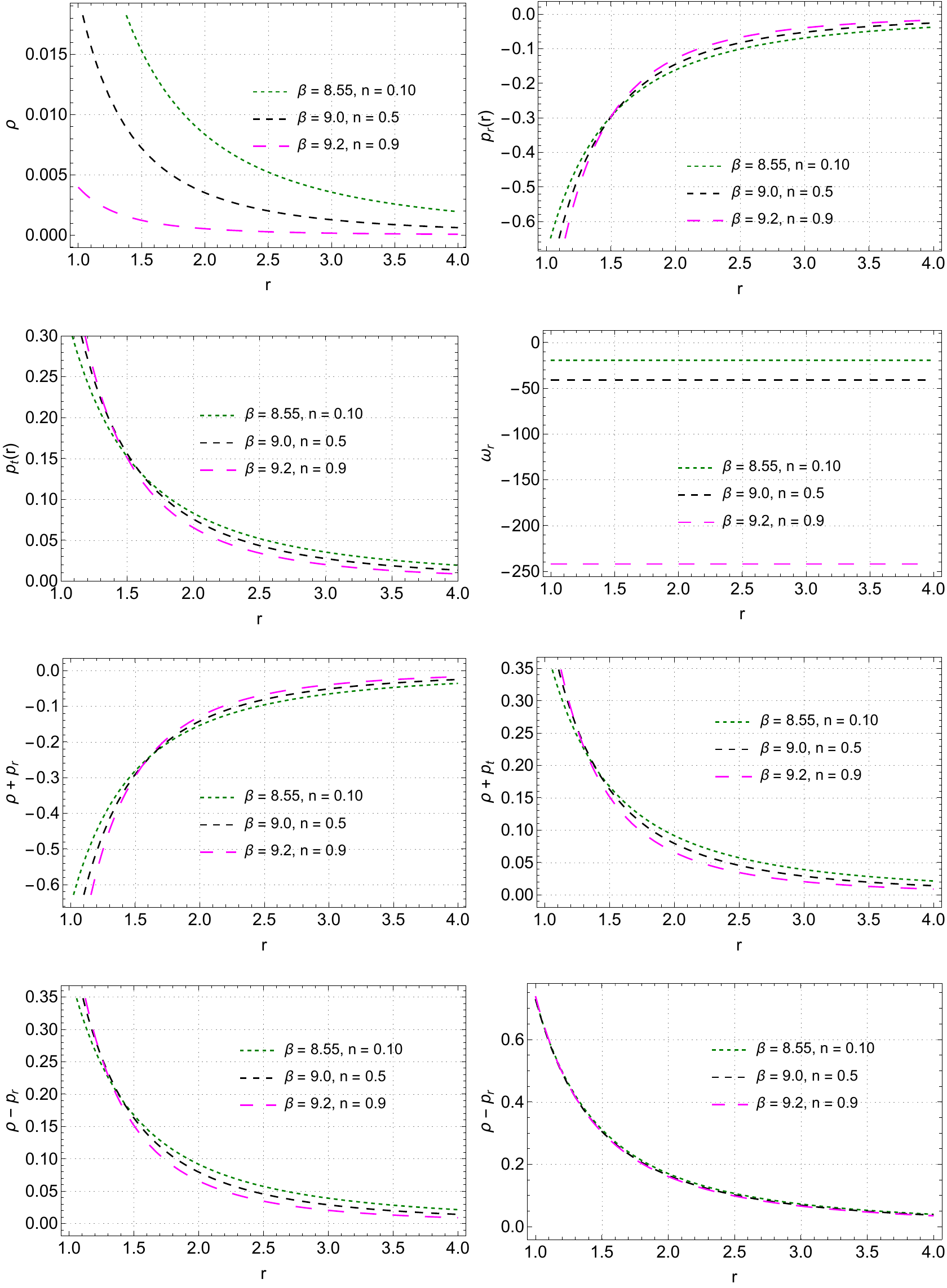}} \caption{The physical parameters and the energy conditions for the shape function \eqref{shape3} is shown for three values of $n$ and $\beta$.}
\label{fig_ec_3}
\end{figure*}
\begin{figure*}[tbh]
\centerline{\includegraphics[scale=0.45]{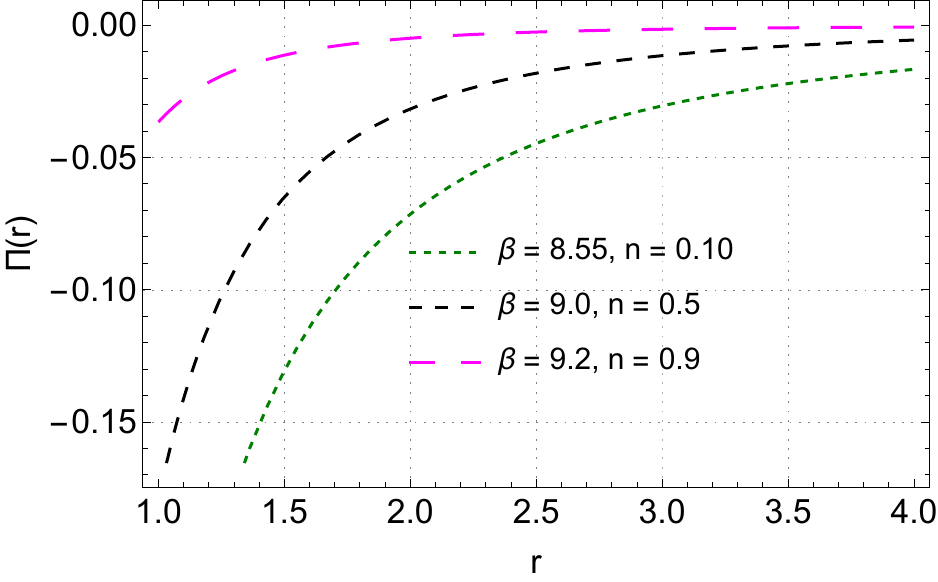} \hspace{0.3cm} \includegraphics[scale=0.45]{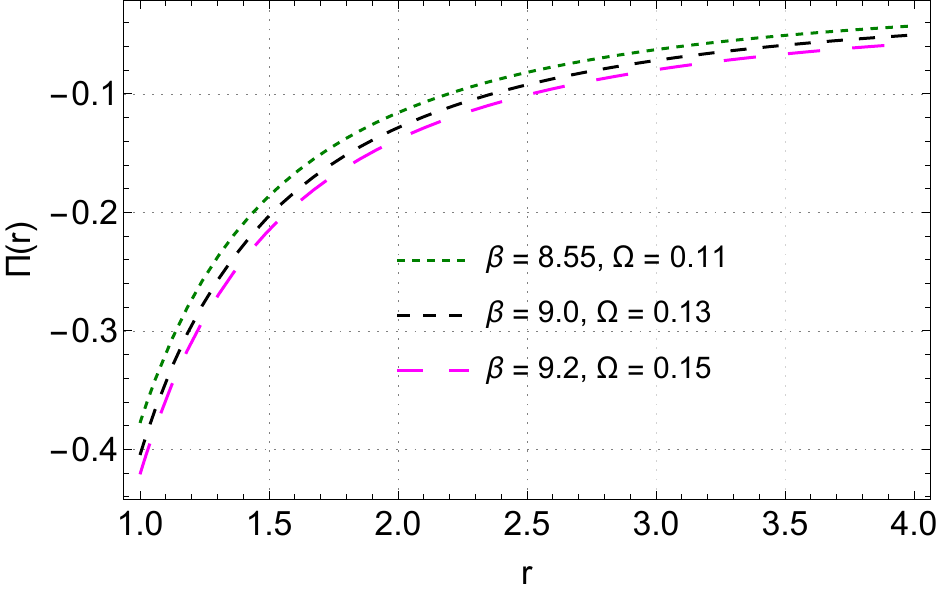}}
\caption{Particle creation pressures with shape functions \eqref{shape3} (left) and \eqref{shape4} (right)}.
\label{fig_pc_2}
\end{figure*}
From Fig. \ref{fig_ec_3}, it is seen that the energy density $\rho$ for this model in the presence of particle creation decreases radially, with maximum value at the throat. However in this case, the rate of decrease of energy density is significantly affected by the variation of $\beta$ and $n$, but the effect of $\beta$ and $n$ on the evolution of the radial pressure $p_r$ is noticeable but not very drastic. Similar to the previous shape functions, it takes on negative values with maximum negative values at the throat which as mentioned earlier, prevents the radial collapse of the wormhole at the throat. However, the tangential pressure remains positive and is significantly sensitive to the values of $\beta$ and $n$. The NEC is violated in terms of radial pressure at the throat but tends to obey as $r \to \infty$, whereas the DEC is obeyed at the throat as well as at $r \to \infty$. In terms of the tangential pressure, the NEC is obeyed at the throat as well as at asymptotic regions but DEC is violated at the throat and obeyed at $r \to \infty$. Also, WEC is violated in terms of the radial pressure but obeyed in terms of the tangential pressure. Moreover, from Eq. \eqref{Pi_3} and Fig. \ref{fig_pc_2} (left), it is seen that the particle creation pressure is negative at the throat to $r \to \infty$. Interestingly, in this case of the shape function \eqref{shape4}, the effective matter governed by the radial EoS parameter $\omega_r = p_r /\rho$ displays a unique behaviour contrasting to the previous cases. Although it behaves like an exotic fluid, it remains constant throughout the entire wormhole.
\subsection{Shape function 3: $b(r) = r \exp [\Omega(r - r_0)]$}
\begin{figure*}[tbh]
\centerline{
\includegraphics[scale=0.45]{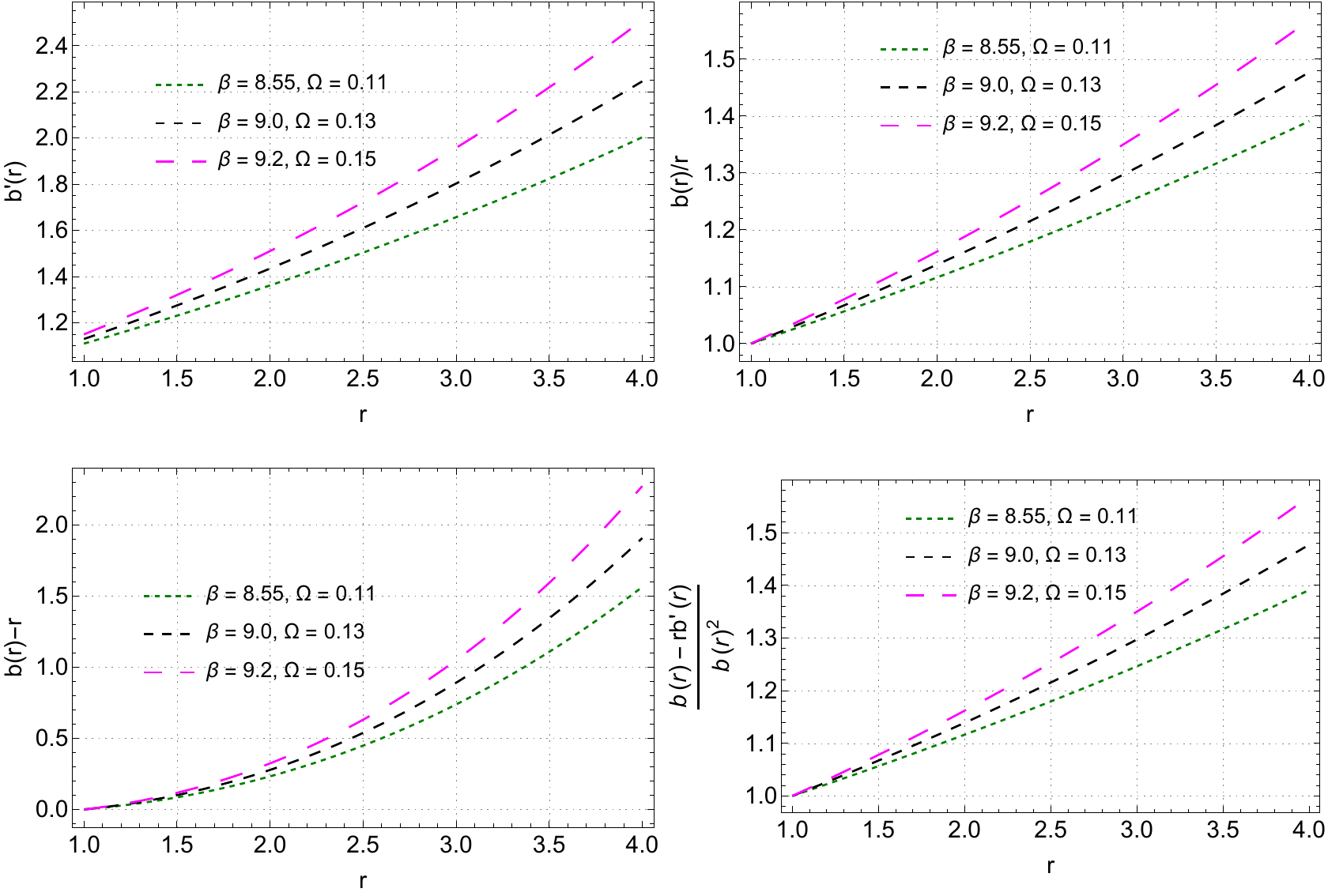}}
\caption{The viability conditions related to the shape function \eqref{shape4} is shown for three values of $n$ and $\beta$.}
\label{cond_sh3}
\end{figure*}

Finally, we consider an exponential-type shape function given by
\begin{equation}
b(r) = r \exp [\Omega(r - r_0)],
\label{shape4}
\end{equation}
where $\Omega$ is the only model parameter and $r_0$ being the throat radius. This type of shape function has been considered widely in Ref. \cite{Samanta2020,Shweta2020,Manjunatha2002} in both GR and modified gravity. In Fig. \ref{cond_sh3}, $b'(r)$, $b(r)/r$, $b(r)-r$ and $\frac{b(r)-r b^{\prime}(r)}{b^2(r)}$ are plotted for different values of $\Omega$, and shows that the validity of the throat, flaring-out and asymptotic flatness conditions are not satisfied.
However for the sake of completeness, let us calculate the physical quantities and analyze the energy conditions. The energy density corresponding to this shape function is
\begin{equation}
\rho = \frac{e^{-(\Omega  (r-r_0))}-r \Omega  e^{-(\Omega  (r- r_0))}}{8 \pi  r^2},
\label{rho4}
\end{equation}
The radial and tangential pressures are obtained as
\begin{equation}
p_r  = \frac{\beta  (\omega +1) \left(e^{-(\Omega  (r- r_0))}-r \Omega  e^{-(\Omega  (r-r_0))}\right)}{8 \pi  r^2}-\frac{e^{-(\Omega  (r- r_0))}}{r^2},
 \label{p4}
\end{equation}
\begin{equation}
 p_t = \frac{e^{\Omega  (r_0-r)} (4 \pi  r \Omega -\beta  (\omega +1) (r \Omega -1))}{8 \pi  r^2},
 \label{pt4}
\end{equation}
In this model, the particle creation pressure can be expressed as
\begin{equation}
\Pi = -\frac{\beta  (\omega +1) \left(e^{-(\Omega  (r-r_0))}-r \Omega  e^{-(\Omega  (r- r_0))}\right)}{8 \pi  r^2}.
\label{Pi_4}
\end{equation}
The quantities associated with the energy conditions are calculated as
\begin{equation}
\rho + p_r  := -\frac{e^{\Omega  (r_0-r)} ((\beta  \omega +\beta +1) (r \Omega -1)+8 \pi )}{8 \pi  r^2},
\label{nec4}
\end{equation}
\begin{equation}
\rho + p_t := \frac{e^{\Omega  (r_0-r)} (-\beta  (\omega +1) (r \Omega -1)+(4 \pi -1) r \Omega +1)}{8 \pi  r^2},
\label{nect4}
\end{equation}
\begin{equation}
\rho - p_r := -\frac{e^{\Omega  (r- r_0)} ((\beta  \omega +\beta -1) (r \Omega +1)+8 \pi )}{8 \pi  r^2},
\label{dec4}
\end{equation}
\begin{equation}
\rho - p_t := \frac{e^{\Omega  (r_0-r)} ((\beta  \omega +\beta -1) (r \Omega -1)+8 \pi )}{8 \pi  r^2},
\label{dect4}
\end{equation}
\begin{figure*}[!th]
\centerline{\includegraphics[scale=0.45]{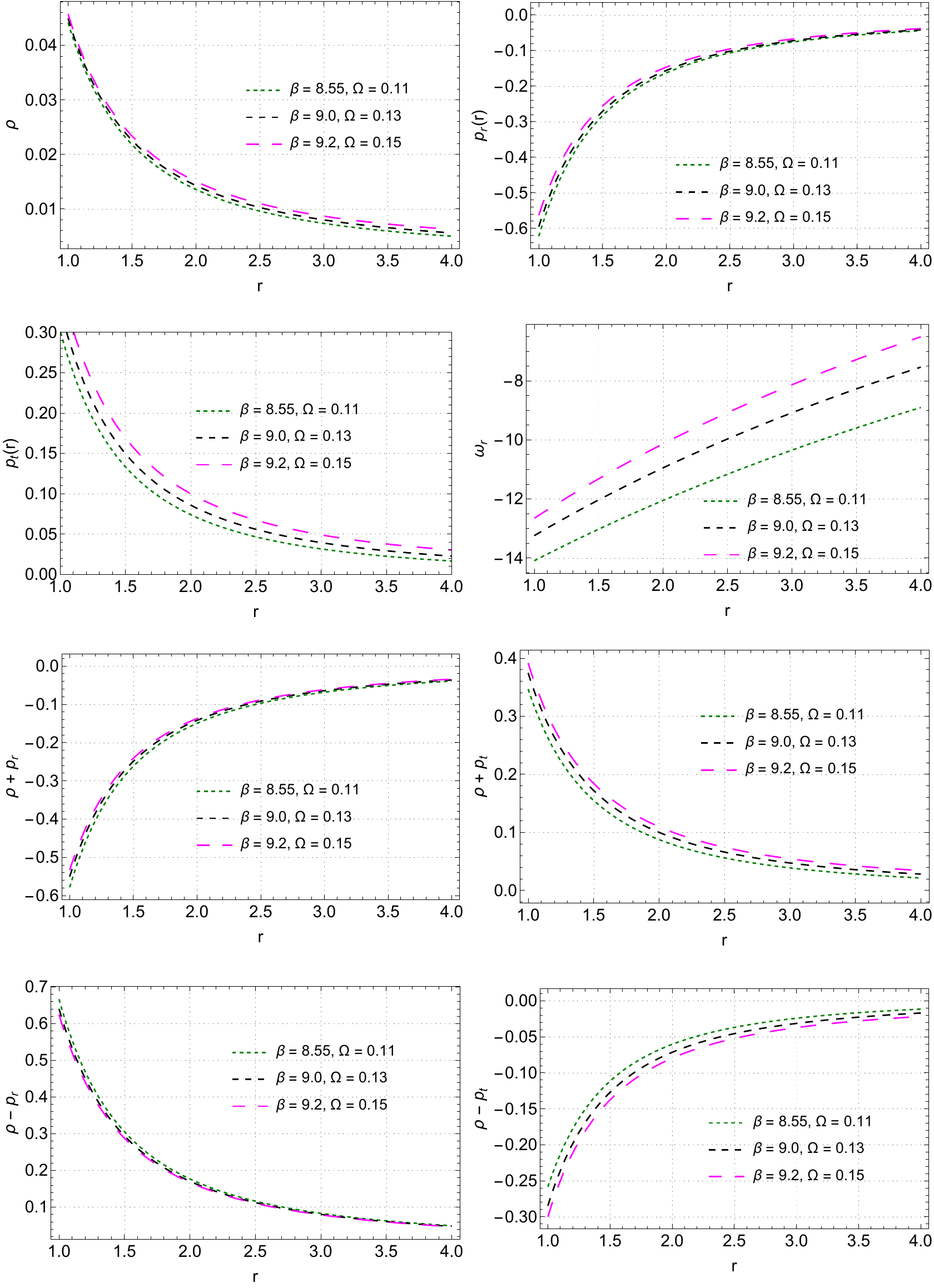}} \caption{The physical parameters and the energy conditions for the shape function \eqref{shape4} is shown for three values of $\Omega$ and $\beta$.}
\label{fig_ec_4}
\end{figure*}
From Fig. \ref{fig_ec_4}, it is seen that the energy density $\rho$ for this model in the presence of particle creation decreases radially, with maximum value at the throat. In this case, the rate of decrease of energy density is very lightly affected by the variation of $\beta$ and $\Omega$. The effect of $\beta$ and $\Omega$ on the evolution of the radial pressure $p_r$ is also seen to be similar but takes on negative values with maximum at the throat. With the same justification provided in the earlier cases, this negative radial pressure plays a role in preventing the radial collapse of the wormhole at the throat. However, the tangential pressure remains positive and is apparently sensitive to the values of $\beta$ and $\Omega$.
We can observe NEC violation in terms of radial pressure at the throat but is obeyed at $r \to \infty$, whereas the DEC is obeyed at the throat as well as at $r \to \infty$. In terms of the tangential pressure, the NEC is satisfied at the throat as well as at asymptotic regions but DEC remains violated at the throat but obeyed at $r \to \infty$. Moreover, WEC is violated for the radial pressure case and satisfied for the tangential pressure one. Also, from Eq. \eqref{Pi_4} and Fig. \ref{fig_pc_2} (right), it is seen that the particle creation pressure is negative at the throat and tends to zero at $r \to \infty$. The particle creation pressure for all the cases takes negative values at the throat which signify that although the perfect fluid matter content obeys the NEC (due to dust like EoS we considered), but the overall violation of NEC is provided by the negative particle creation pressure at the throat. Finally, the radial EoS parameter $\omega_r = p_r /\rho$, is found to evolve with negative values, that shows exotic matter like behaviour at the throat but takes on higher negative values away from the throat. Although the physical parameters seems to behave well, the overall structure of a wormhole with the shape function \eqref{shape4} is not allowed in the framework of particle creation, as the viability conditions are broken as discussed above.

\subsection{Matching Conditions of Derived Shape Function with the Toy Shape Functions: Comparative Analysis}
Let us now make a comparison of the shape function \eqref{non-lin_shape}, derived from the physical judgements of particle creation with the toy shape functions \eqref{shape2}, \eqref{shape3} and \eqref{shape4}. If we need to estimate the explicit expression for $\alpha$ and $K$ from the shape function \eqref{non-lin_shape}, with reference to the toy  models \eqref{shape2}, \eqref{shape3} and \eqref{shape4}, we may equate the proposed general form of particle creation pressure given by Eq. \eqref{pc2_powlaw} with the derived particle creation pressures for each toy model, and then solve for $\alpha$ and $K$. The resulting expressions for $\alpha$ and $K$ will then determine the conditions under which the toy shape functions will exactly mimic our derived shape function \eqref{non-lin_shape}. The estimated general forms for $\alpha$ and $K$ corresponding to each toy model are tabulated in Tab. \ref{tab_match}. 

\begin{table*}[!h]
\centering
\begin{tabular}{||c|c|c|c||}
\hline
$b(r)$ & $\displaystyle \frac{r_0 \tanh (\epsilon r)}{\tanh (\epsilon r_0)}$ & $\displaystyle r_0^n r^{1-n}$ & $\displaystyle r \exp \left( (r - r_0) \Omega \right)$ \\
\hline
$\alpha$ & $\displaystyle \frac{\log (8 \pi ) - \log \left( -\frac{\beta r_0 (\omega + 1) \epsilon \coth (r_0 \epsilon) \text{sech}^2 (r \epsilon)}{k r^2} \right)}{\log (r)}$ & $\displaystyle \frac{\log (8 \pi ) - \log \left( \frac{\beta (n-1) (\omega + 1) r^{-n-2} r_0^n}{k} \right)}{\log (r)}$ & $\displaystyle -\frac{\log \left( -\frac{\beta (\omega + 1) e^{(r - r_0) \Omega} (r \Omega + 1)}{8 \pi k r^2} \right)}{\log (r)}$ \\
\hline
$K$ & $\displaystyle -\frac{\beta r_0 (\omega + 1) \epsilon r^{\alpha - 2} \text{sech}^2 (r \epsilon) \coth (r_0 \epsilon)}{8 \pi}$ & $\displaystyle \frac{\beta (n-1) (\omega + 1) r_0^n r^{\alpha - n - 2}}{8 \pi}$ & $\displaystyle -\frac{\beta (\omega + 1) r^{\alpha - 2} (r \Omega + 1) e^{\Omega (r - r_0)}}{8 \pi}$ \\
\hline
\end{tabular}
\caption{Estimated general expressions for $\alpha$ and $K$ in the derived shape function \eqref{non-lin_shape} after matching with the toy shape functions.}
\label{tab_match}
\end{table*}
\section{Conclusion}
\label{conc}
The concept of gravitational particle creation phenomena of open thermodynamic cosmological systems, was proposed as a potential explanation to the late-time accelerated expansion of the Universe and inflationary expansion during the early stages of the Universe. This mechanism involves the generation of repulsive gravitational effects, resulting in negative pressure, which could explain the accelerated expansion of the Universe in its later stages. Gravitational particle creation allows for speculation about violations of the NEC, which is crucial for the maintenance of a wormhole structure. This study explores the hypothesis that the geometry of a wormhole can be supported by the negative pressure created by particles. The initial section of the paper focuses on deriving the shape function of a wormhole based on a specific form of particle creation pressure. Specifically, we proposed an inverse-powerlaw form of particle creation pressure with the radial coordinate, with a source function that is linear in the Hubble parameter at the current time, denoted by $\Gamma = 3\beta H_0$ with $\beta$ as a free parameter. Through out the paper, owing to the finiteness of the redshift function, we considered a constant redshift function for the analysis. The shape function obtained seems to adhere to the required criteria for a viable wormhole structure and violates the NEC with a positive energy density. Also the 2D and 3D embedding for the derived wormhole solution is also carried out. In the later part of the paper, in the context of particle creation mechanism we obtained the radial variation of energy density, radial and tangential pressure, and the particle creation pressure within the wormhole geometry, by assuming a few well known shape functions given by Eqs. \eqref{shape2}, \eqref{shape3} and \eqref{shape4}. The radial NEC is found to be violated at the throat in the first two case, which resonates with the behaviour of so-called \emph{exotic matter} but without the need of introducing explicit form of exotic matter density like chaplygin gas \cite{Chakraborty2009Mar,Lobo2006Mar,Elizalde2018Dec,Eiroa2009Aug,
Eid2018Oct,Sharif2013Sep,Gorini2008Sep,Bhatti2023Dec} or phantom matter \cite{Jamil2010Jun,Jamil2009Feb,Sahoo2018Jul,Sahoo2019Aug,Lobo2005Jun,
Lobo2005Apr,Gonzalez2009Mar,Lobo2013Apr,Sushkov2005Feb}. But in the third case, the viability conditions are violated, which therefore can be ruled out as a viable shape function in the framework of particle creation. Finally, we obtained the possible general forms of $\alpha$ and $K$, the parameters those dictate the role of particle creation pressure. These general expressions represents the conditions under which the derived wormhole shape function shall mimic the assumed toy shape functions.

At this point, we may compare our study with the one carried out by  Pan and Chakraborty \cite{Pan2015Jan} where they have studied dynamical evolving wormholes in the background of thermodynamical particle creation. They showed that evolving wormhole solutions can be obtained when the nature of the perfect fluid is phantom in nature. Contrastingly, we have found that even in the presence of a normal fluid, wormhole formation is possible.

This study explores the possibility for developing a viable wormhole geometry supported by the back reaction pressure generated from the irreversible particle creation, within an open thermodynamic system. Nevertheless, our work does not address the effect of particle creation on the timelike or null geodesics within the wormhole structure or related aspects, as it is beyond the current study's scope. This prospect will be further investigated in the future work.
\bibliography{ref_wh.bib}
\end{document}